\documentclass[aps,pra,preprintnumbers,twocolumn,groupedaddress,showpacs,amsmath,amssymb]{revtex4}

\usepackage{graphicx}
\usepackage{amsmath}
\usepackage{bm}

\def\6#1{{\underline{#1}}}
\def\m6#1{{\underline{#1}\,}}

\newdimen\Tdim
\def\ispan{{\setbox0=\hbox{i}%
\Tdim\ht0\advance\Tdim\dp0\rule[-\dp0]{0pt}{\Tdim}}}
\def\jspan{{\setbox0=\hbox{j}%
\Tdim\ht0\advance\Tdim\dp0\rule[-\dp0]{0pt}{\Tdim}}}
\def\Tspan#1{{\setbox0=\hbox{#1}%
\Tdim\ht0\advance\Tdim\dp0\advance\Tdim.55ex\rule[-\dp0]{0pt}{\Tdim}\box0}}

\def\be{\begin{eqnarray}}
\def\ben{\begin{eqnarray*}}
\def\ee{\end{eqnarray}}
\def\een{\end{eqnarray*}}

\def\=:{=\hspace{-.7em}\raisebox{1.1ex}{.}\hspace{.1em}\raisebox{-0.2ex}{.} }

\newcommand {\beq}{\begin{eqnarray}}
\newcommand {\eeq}{\end{eqnarray}}
\newcommand {\non}{\nonumber\\}

\begin{document}
\preprint{YGHP-11-46}

\title{Vortex trimer in three-component Bose-Einstein condensates}


\author{Minoru Eto$^{1}$}
\author{Muneto Nitta$^{2}$}
\affiliation{
$^1$Department of Physics, Yamagata University, Yamagata 990-8560, Japan \\
$^2$Department of Physics, and Research and Education Center for Natural 
Sciences, Keio University, Hiyoshi 4-1-1, Yokohama, Kanagawa 223-8521, Japan}


\date{\today}

\begin{abstract}
Vortex trimer is predicted in three-component Bose-Einstein condensates (BEC's) 
with internal coherent couplings.
The molecule is made by three constituent vortices which are bounded by domain walls of the relative phases.
We show that the shape and the size of the molecule can be 
controlled by changing the internal coherent couplings.

\end{abstract}


\maketitle

\section{Introduction}

Recent advances in realizing Bose-Einstein condensates (BEC's) in ultracold atomic gases 
have opened new possibilities of quantum physics \cite{Pethickbook,Ueda}.
One of them is interpenetrating superfluids, 
a mixture of two or more superfluids. 
Such multicomponent BEC's can be realized when 
more than one hyperfine spin state is simultaneously populated 
or when more than one species of atoms are mixed.
The $s$-wave scattering wave-length can be tuned via a Feshbach resonance 
\cite{Thalhammer,Papp,Tojo}. 
Moreover, recent experimental achievement of a condensate of ytterbium offers 
condensations up to five components \cite{Fukuhara:2007}.
Stability condition of multicomponent BEC's was 
studied in Ref.~\cite{Roberts:2006}. 
One of the most important consequences of superfluidity is 
the existence of vortices. 
Vortices in multicomponent BEC's have been realized experimentally 
\cite{Matthews,Schweikhard:2004}, 
and structures of those vortices are much  
richer than those of single components 
\cite{
Kasamatsu2,Eto:2011,Aftalion:2011}. 

In the case of multiple hyperfine spin states, 
the internal coherent coupling 
between multiple components can be introduced by 
Rabi oscillations.
This case is similar to two gap superconductors 
with Josephson coupling between the two gaps.
A sine-Gordon domain wall of a phase difference of two components 
is allowed \cite{Son:2001td}.
Moreover, an integer vortex is split into 
two fractional vortices with fractional circulations, 
and they are connected by a sine-Gordon domain wall 
with the total configuration being 
a molecule of two constituent vortices, 
namely, a vortex dimer \cite{Kasamatsu:2004,Turner:2009}.
Therefore it is natural to ask whether 
a molecule made of more than two vortices 
is possible in some case, 
or how domain walls connect among them if it is possible.

In this paper we explicitly construct a vortex trimer, 
namely, a molecule made of three constituent vortices 
winding around respective three components of BEC's with 
internal coherent couplings induced by Rabi oscillations.
Varying the internal coherent couplings, 
the shape of the molecule is changed accordingly. 
We also find a dependence of the size of the vortex trimer 
on the magnitude of the Rabi frequency.


\section{The Gross-Pitaevskii model}

We consider three-component BEC's of atoms with equal mass $m$, 
described by
the condensate wave functions $\psi_i$ ($i=1,2,3$) with the energy functional
\beq
E&=& \sum_{i,j}\int d^2x 
 \bigg(-\frac{\hbar^2}{2m}\psi_i^*\nabla^2\psi_i \delta_{ij}
+ \frac{g_{ij}}{2}|\psi_i|^2 |\psi_j|^2 \non
&&- 
\mu_i|\psi_i|^2\delta_{ij} - \omega_{ij}\psi_i^*\psi_j
\bigg),
\label{eq:gp}
\eeq
where atom-atom interactions are characterized by the coupling constants $g_{ij} = g_{ji}$,
$\mu_i$ is a chemical potential, and a symmetric tensor $\omega_{ij} = \omega_{ji}$ ($\omega_{ii} = 0$)
stands for the Rabi frequency between the $i$th and $j$th components.
In this paper, we consider the case with $\mu_1=\mu_2=\mu_3 \equiv \mu$, $g_{11}=g_{22}=g_{33} \equiv g$, and
$g_{12} = g_{23} = g_{31} \equiv \tilde g$ for simplicity, 
but the general case is straightforward.
We also assume $g + 2 \tilde g > 0$ for the stability of ground states.

In the following, we will separately study two cases: the case with
$g\neq \tilde g$ ($\det g_{ij} \neq 0$) and the $U(3)$ symmetric case with $g=\tilde g$
($\det g_{ij} = 0$).

\begin{figure}[t]
\begin{center}
\begin{tabular}{cc}
\includegraphics[width=4.5cm]{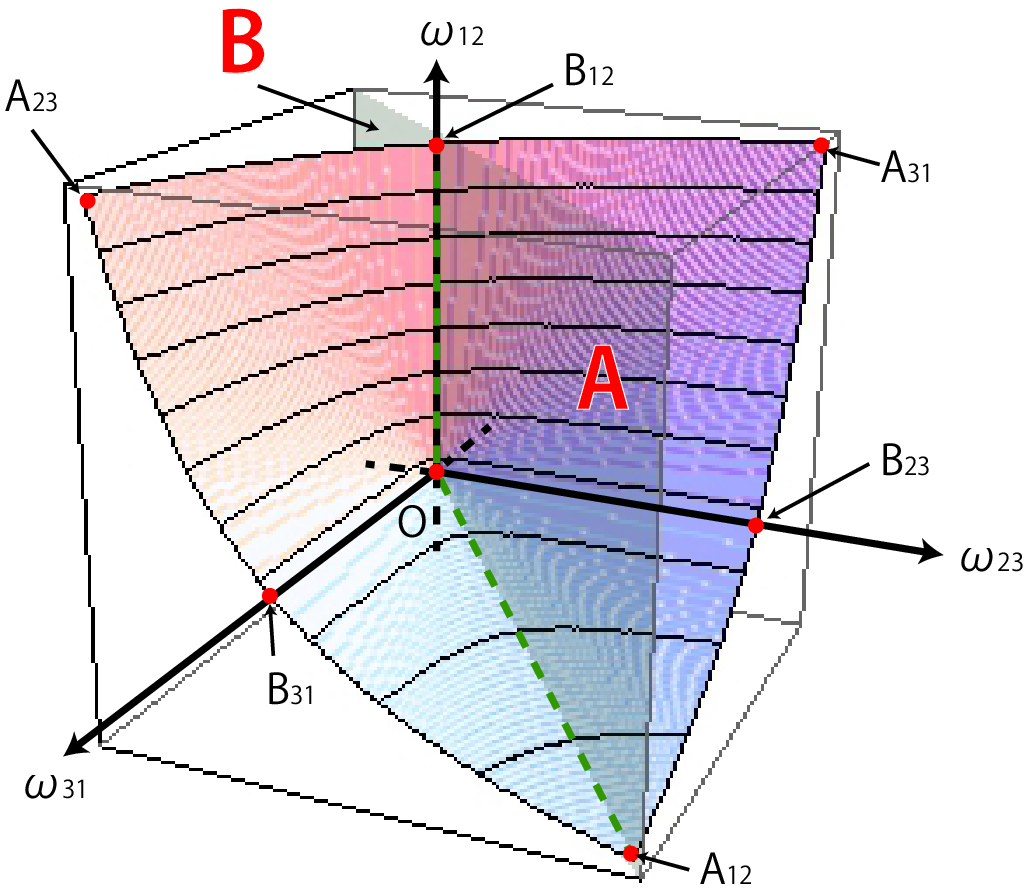}
\includegraphics[width=4cm]{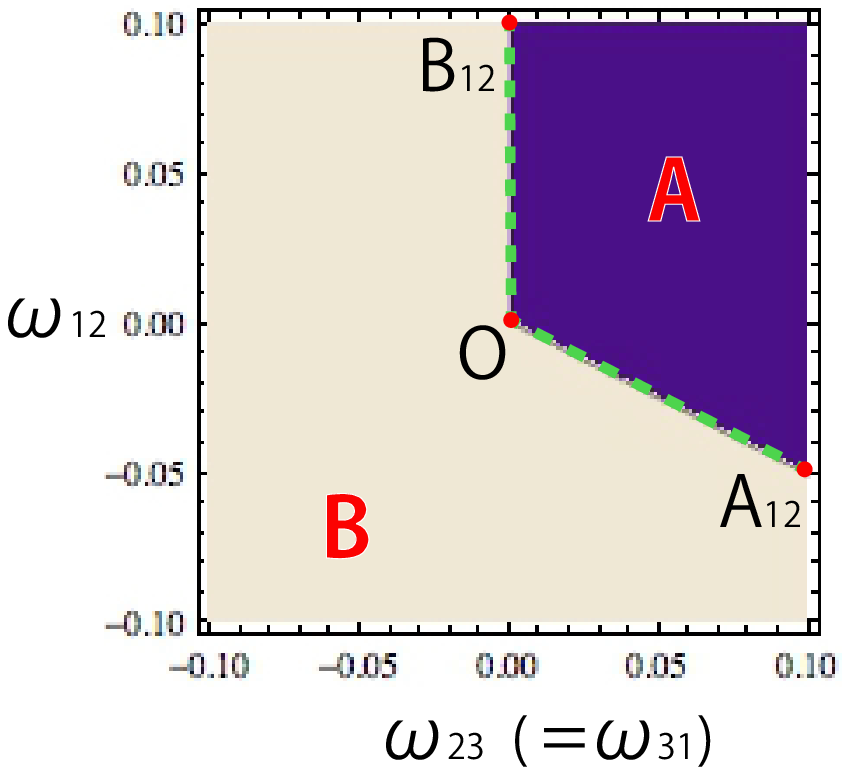}
\end{tabular}
\caption{(Color online)
The parameter region 
where Eq.~(\ref{eq:true_cond}) holds.
The left panel shows the boundary surface on 
which Eq.~(\ref{eq:true_cond}) holds inside (the region A) 
while it does not hold outside (the region B).
The plot range of the left panel is $-0.05 \le \omega_{ij} \le 0.1$.
The right panel shows the cross section $\omega_{23} = \omega_{31}$.
Some points $\{A_{ij},B_{ij},O\}$ on the surface are shown: 
$A_{12}$ is placed at $(\omega_{12},\omega_{23},\omega_{31})=(-0.05,0.1,0.1)$, 
$B_{12}$ is at $(0.1,0,0)$, and the other points are obtained by 
the $\mathbb{Z}_3$ rotations around $O$.
}
\label{fig:region}
\end{center}
\end{figure}

Let us first study the former case.
When all the Rabi frequencies vanish, the ground state is given by
\beq
|\psi_i|^2 = v^2,\quad v \equiv \sqrt{\frac{\mu}{g+2\tilde g}},\qquad (i=1,2,3).
\eeq
The topology of the ground state is characterized $\pi_1[U(1)^3] = \mathbb{Z} \oplus \mathbb{Z} \oplus \mathbb{Z}$.
Once the small Rabi frequencies are turned on, 
only the overall $U(1)$ symmetry remains contact and
the homotopy group also reduces to $\pi_1[U(1)] = \mathbb{Z}$.
At the same time, the magnitudes of the condensates are modified. 
This is because the Rabi frequencies yield potentials on the 
relative phases of $\theta_i = \arg \psi_i$. The ground state can be obtained by solving
a variational equation $\delta E/ \delta \psi_i = 0$. We denote the condensate wave function of
the ground state by
\beq
\psi_i = v_i e^{i\theta_i},\quad (v_i > 0).
\eeq
The last term in Eq.~(\ref{eq:gp}) 
can be written as $- 2v_iv_j\omega_{ij}\cos(\theta_i - \theta_j)$.
For $\omega_{ij} > 0$, 
the phases $\theta_i$ and $\theta_j$ tend to coincide, $\theta_i = \theta_j$, 
to reduce this interaction energy. 
Therefore, when $\omega_{ij} > 0$ hold for all $i,j=1,2,3$, 
all the phases coincide, 
\beq
\theta_1 = \theta_2 = \theta_3,
\label{eq:true_cond}
\eeq
in the ground state. 
We numerically find that relation (\ref{eq:true_cond}) is satisfied in the parameter region A 
shown in  Fig.~\ref{fig:region}.
We note that Eq.~(\ref{eq:true_cond}) holds 
not only when all $\omega_{ij}$'s are positive
but also in the region where one of the $\omega_{ij}$'s is negative.
In this paper, we consider region A, 
where relation (\ref{eq:true_cond}) holds. 
The other region is frustrated, which will be studied elsewhere.

\section{Vortex configurations}

The nontrivial first homotopy group immediately 
leads to the existence 
of superfluid vortices. In particular, the case with $g \neq \tilde g$ would have three different kinds of vortices
because of the homotopy group $\pi_1[U(1)^3] = \mathbb{Z} \oplus \mathbb{Z} \oplus \mathbb{Z}$ 
(when $\omega_{ij} = 0$).

Let us consider an integer vortex configuration in which all the condensations $\psi_i$ have unit winding in
$U(1)$'s. The asymptotic behavior 
of such a configuration 
at large distance from the vortices
should be
\beq
(\psi_1,\psi_2,\psi_3) 
\to (v_1e^{i\theta},v_2 e^{i\theta},v_3e^{i\theta}),
\label{eq:asym_true}
\eeq
which satisfies the ground-state condition (\ref{eq:true_cond}).
Here $\theta$ stands for the angular coordinate as $x + i y = r e^{i\theta}$.
We will show that this vortex is deformed to a vortex trimer made of three constituent vortices,
$(v_1e^{i\theta}, v_2 ,v_3)$,
$(v_1,v_2e^{i\theta} ,v_3)$, and
$(v_1,v_2,v_3e^{i\theta})$, which
we call (1,0,0)-, (0,1,0)- and
(0,0,1)-vortices, respectively.
Since it is a dynamical problem if the constituent vortices make a bound state or not,
let us see the two cases $\omega_{ij}=0$ and $\omega_{ij}\neq 0$  separately.
 
\begin{figure}[t]
\begin{center}
\includegraphics[width=9cm]{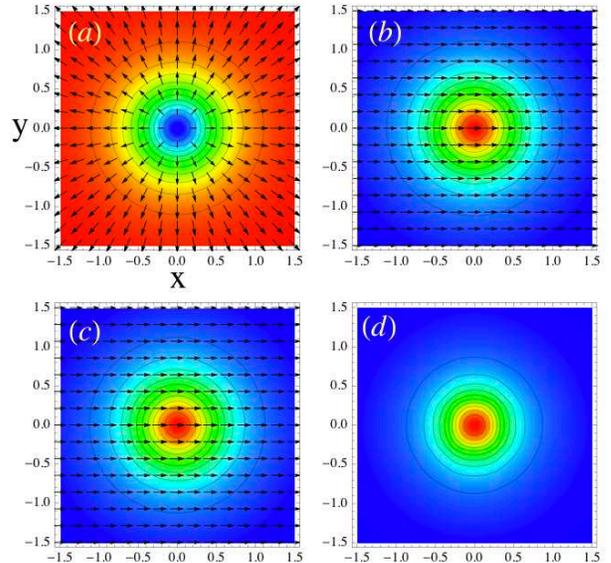}
\caption{(Color online) The panels (a), (b), and (c) show the profiles of the density $|\psi_1|^2$, $|\psi_2|^2$, and $|\psi_3|^2$
for the $(1,0,0)$-vortex in the case where $g\neq\tilde g$,
respectively. 
The arrows show a phase vector $({\rm Re}(\psi_i), {\rm Im}(\psi_i))$.
The energy density is shown in the panel (d).
We choose $\hbar=m=1,g = 1000, \tilde g=900, \mu=100$. 
The Rabi frequencies are $\omega_{12} = \omega_{23} = \omega_{31} = 0$.
The sizes of the boundary are $L=-15$ to $L=15$ for both $x$ and $y$.}
\label{fig:1c}
\end{center}
\end{figure}

When the Rabi frequencies vanish ($\omega_{ij}=0$), the constituent vortices do not make a molecule
since they repel each other.
Instead, the constituent vortex can exist alone; see Fig.~\ref{fig:1c}, where
a numerical solution \footnote{We use the relaxation (the imaginary time) method for the numerical computation in this paper.
We choose the boundary condition with the fixed phase winding and the constant density as given 
in Eq.~(\ref{eq:asym_true}).} 
of the constituent vortex is shown.
The tension (energy per unit length) of the constituent vortex is given by
$\frac{\pi \hbar^2 v^2}{m}\ln \frac{L}{\xi}$
with system size $L$ and healing length $\xi$.

On the other hand, when the Rabi frequencies are not zero, 
the unit constituent vortex alone is unstable because semi-infinite domain walls are attached to it.
This domain wall supplies attractive force between the constituent vortices and can be balanced
with repulsion among them, so that the constituent
vortices form a vortex trimer 
(vortex dimers are stable only when 
two of the Rabi frequencies are zero).
To understand this better, it is useful to consider a reduced model from Eq.~(\ref{eq:gp}) by 
fixing the amplitudes $|\psi_i| \simeq v_i$ ($v_i \simeq v$ for simplicity). 
Then we are left with three phases:
\beq
\Theta = \sum_i \theta_i,\quad \delta_{1,2,3} = \theta_{2,3,1} - \theta_{3,2,1}.
\eeq
The Hamiltonian of the reduced model is given by
\beq
H = \frac{\hbar^2 v^2}{6m}\left[(\nabla \Theta)^2 + \sum_i\left((\nabla \delta_i)^2 
- \tilde \omega_i \cos\delta_i \right) \right],
\label{eq:sine-Gordon}
\eeq
where we have introduced the renormalized couplings
$\tilde \omega_{1,2,3} =  \frac{12m}{\hbar^2}\omega_{23,31,12}$. 
This approximation is valid
only when the Rabi frequencies are much smaller than the other coupling constants. \footnote{
This condition meets our purpose, since we would not like to consider too large Rabi frequencies such that 
the size of the molecule given in Eq.~(\ref{eq:size}) becomes comparable to the healing length of the system.
In this case, 
the vortex trimer cannot be distinguished from an integer vortex.
} 
For example, let us consider the $(1,0,0)$-vortex, 
with relative phases given by
\beq
\delta_1 = 0,\quad \delta_2 = -\theta_1,\quad \delta_3=\theta_1.
\eeq
Then the potential term reads
\beq
V= - \frac{\hbar^2v^2}{6m}\left(\tilde\omega_2\cos\delta_2  + \tilde\omega_3 \cos\delta_3\right).
\eeq
When $\tilde\omega_{2,3} > 0$, $\delta_2 = \pi$ is an unstable point and
a semi-infinite domain wall appears on the negative region of the real axis.
Its tension  is given by
\beq
T_1 
= \sqrt{T_{12}^2 + T_{31}^2},\quad
T_{ij} = \frac{8\sqrt{6}}{3}\frac{\mu\hbar  \sqrt{\omega_{ij}}}{\sqrt{m}(g+2\tilde g)}.
\label{eq:string_tension}
\eeq
This is the origin of the attractive force between the constituent vortices.
Note that, since we have two relative phases $\delta_2$ and $\delta_3$, 
one may naturally imagine two independent domain walls. 
Each domain wall has the tension $T_{31}$ (when we set $\omega_{12}=0$)
and $T_{12}$ (when we set $\omega_{31}=0$).
However, for the (1,0,0)-vortex, the two 
relative phases are  related as $\delta_2 = - \delta_3$ and these two domain walls stick together and form a bound state.
Indeed, the total tension  $T_1$ is the square root of the sum of $T_{12}^2$ and $T_{31}^2$ as shown in Eq.~(\ref{eq:string_tension}),
which is smaller than the sum of  the two tensions, $T_1 \le T_{12} + T_{31}$.

\begin{figure*}
\begin{center}
\includegraphics[width=15.5cm]{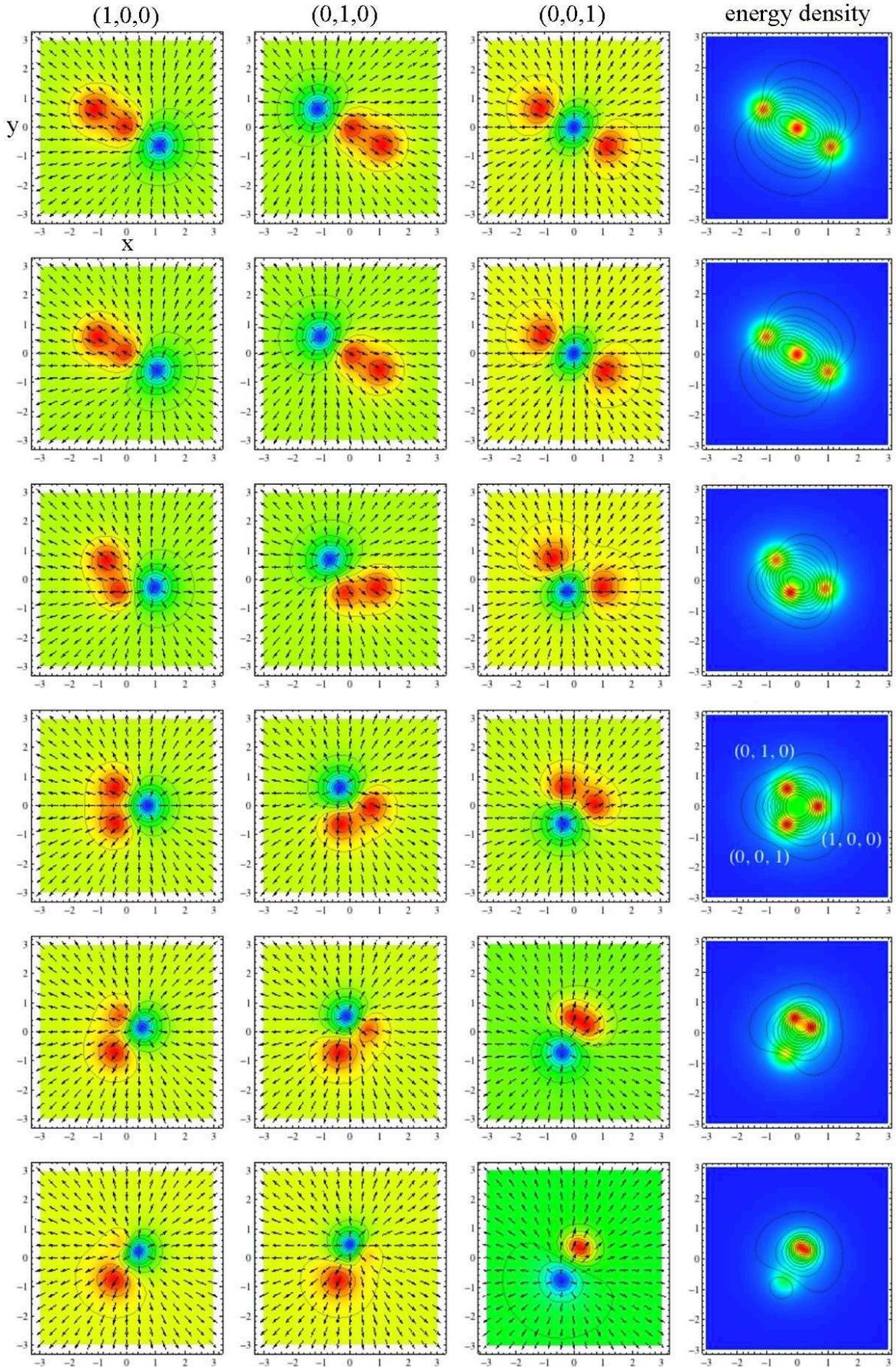}
\caption{(Color online) The left three panels show the profiles of the density $|\psi_i|^2$
and the phases $({\rm Re}[\psi_i],{\rm Im}[\psi]_i)$
for the unit vortex trimer in the case $g \neq \tilde g$,
respectively. The rightmost panel shows energy density and the contour corresponds to the Rabi potential.
The constants are taken as $\hbar = m = 1$, $\mu=100$, $g=1000$, $\tilde g = 900$, and 
$\omega_{23} = \omega_{31} = 0.05$. 
We change the Rabi frequency $\omega_{12}$ from the top  to the bottom as
$\omega_{12}=-0.01, 0, 
0.01,
0.05, 
0.2, 0.5$, respectively.
The sizes of the boundary are $L=-8$ to $L=8$ for both $x$ and $y$.
}
\label{fig:mol1}
\end{center}
\end{figure*}

\section{Numerical solutions of vortex trimers}

We have numerically found stable vortex trimers 
as unique solutions under the boundary condition (\ref{eq:asym_true}) 
in a wide range of the parameter region A
in Fig.~\ref{fig:region}.
Here in Fig.~\ref{fig:mol1} we
show several numerical solutions as examples
with $g=1000$ and $\tilde g = 900$ 
($\mu=100$ and $m=\hbar=1$).
The rightmost panels show the energy density in which the partonic structure is clearly seen.
The contours therein show contributions from the last term of Eq.~(\ref{eq:gp}).
Since the distance between the constituent vortices are close, 
we cannot see domain walls.
Nevertheless, qualitative estimation from the reduced model is quite useful, as will be seen below.

Each line of Fig.~\ref{fig:mol1} gives a molecule with 
different Rabi frequencies. 
First of all, the fourth line of Fig.~\ref{fig:mol1} shows a $\mathbb{Z}_3$ symmetric trimer where
the Rabi frequencies are all equal as $\omega_{12} = \omega_{23} = \omega_{31} = 0.05$ .
One can see that the phases at the spatial infinity are indeed aligned ($\theta_1=\theta_2=\theta_3$).
Next, by changing $\omega_{12}$ from the symmetric case, 
we can observe how the shape of the trimer is deformed.
Since the Rabi frequency $\omega_{12}$ controls the interaction between the (1,0,0)- and the
(0,1,0)-vortices, the equilateral triangle is deformed to an isosceles triangle.
The third line of Fig.~\ref{fig:mol1} shows the vortex trimers with
$\omega_{12}=0.01$, in which
the attractive force between the (1,0,0)- and the (0,1,0)-vortices
is smaller than those between the other two pairs. Therefore the internal angle at the vertex at the  (0,0,1)-vortex
is larger than $\pi/3$. We also show the vortex trimer when $\omega_{12} = 0$ in the second line of Fig.~\ref{fig:mol1}.
Since no attractive force exists between the (1,0,0)- and the (0,1,0)-vortices, the shape of the molecule becomes a stick
as expected.
We also find the molecule even when $\omega_{12}$ is negative $(= -0.01)$ while $\omega_{23} = \omega_{31} = 0.05$; see the first line of
Fig.~\ref{fig:mol1}. 
We still observe a stick-type molecule whose length is slightly larger than that for $\omega_{12} = 0$.
\begin{figure}[t]
\begin{center}
\includegraphics[width=8cm]{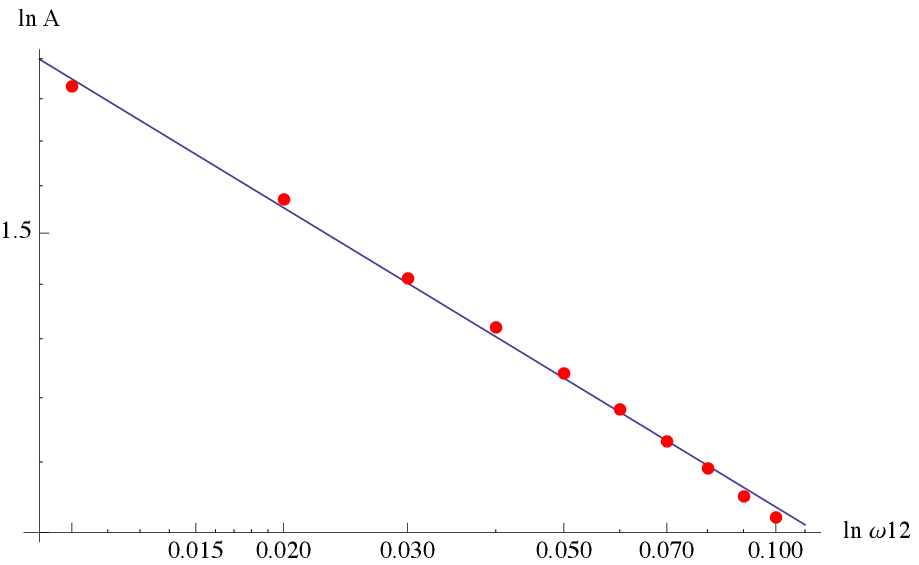}
\caption{(Color online) Loglog-plot of $|\omega_{12}|$ vs the length ($A$) of an edge of the equilateral triangle with
$\omega_{12} = \omega_{23} = \omega_{31}$. The corresponding configuration is given in the fourth line
of Fig.~\ref{fig:mol1}. The parameters are fixed as $\hbar = m = 1$, $\mu=100$ and $g=1000$ and $\tilde g = 900$. }
\label{fig:size_fit}
\end{center}
\end{figure}
In the last two lines of Fig.~\ref{fig:mol1}, we have chosen $\omega_{12} = 0.2$ and $0.5$, which are larger 
than $\omega_{23} = \omega_{31} = 0.05$ 
(the $\mathbb{Z}_3$-symmetric case). 
Since the attractive force between the (1,0,0)- and the (0,1,0)-vortices
is stronger than those for the other two pairs, we see that the corresponding edge of the triangle becomes shorter
than the other two edges. Since $\omega_{12} = 0.5$ yields
an attractive force that is too strong, the triangle collapses, as shown in the last line of Fig.~\ref{fig:mol1}.

We have seen that the shape of the triangle changes according to the choice of the Rabi frequencies.
Here, we investigate a correlation of the Rabi frequencies, set to be equal 
$\omega_{12} = \omega_{23} = \omega_{31} \equiv \omega$, and the size of the equilateral triangle,
see Fig.~\ref{fig:size_fit}.
We numerically find the relation
\beq
A \simeq  0.56 ~\omega^{-0.25},
\label{eq:size}
\eeq
for the range $0.01 \le \omega \le  0.1$, where $A$ stands for the length of the edge of the
equilateral triangle.

\begin{figure}[h]
\begin{center}
\includegraphics[width=8.5cm]{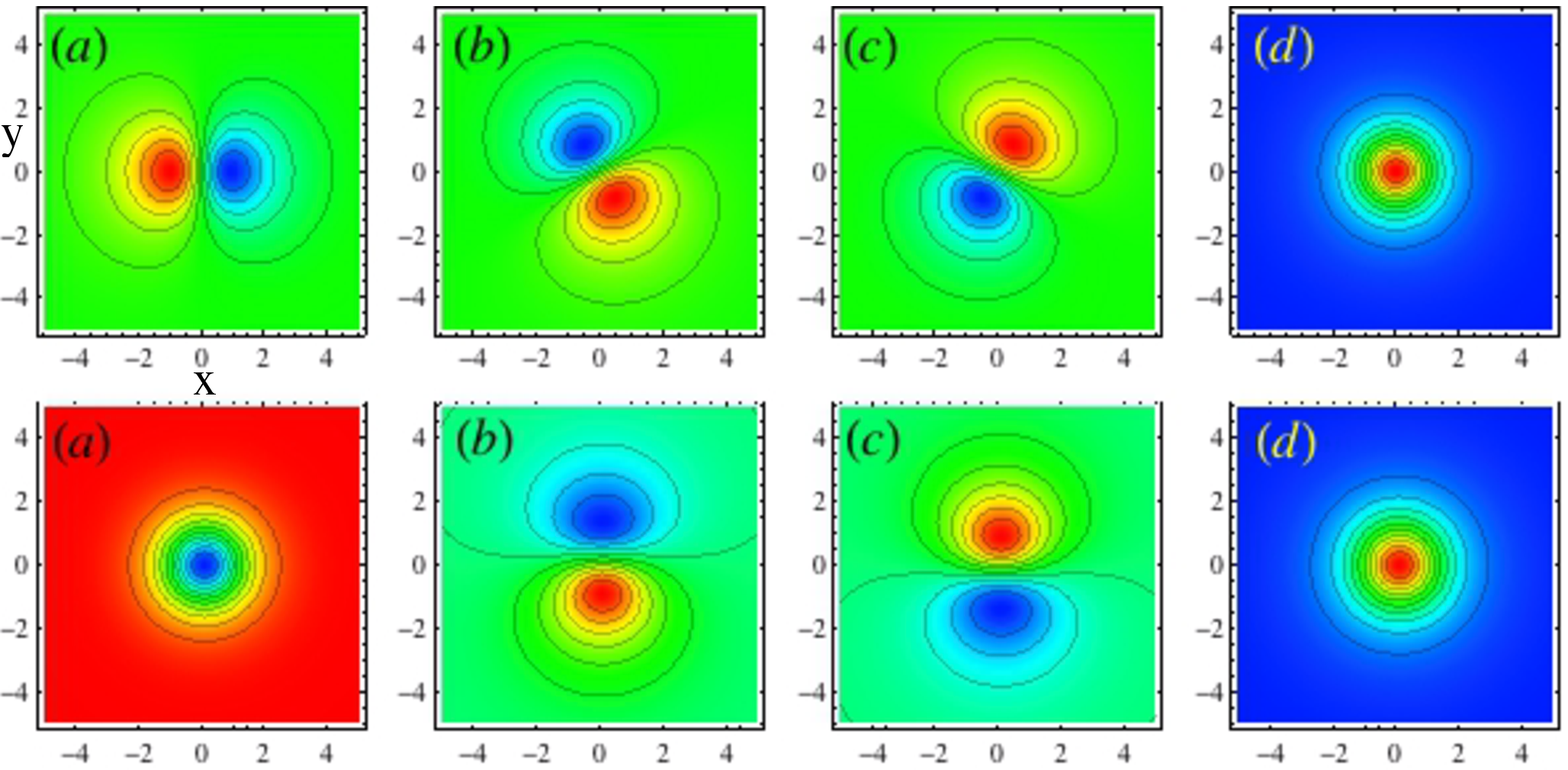}
\caption{(Color online) The panels (a), (b), and (c) show the profiles of the density $|\psi_1|^2$, $|\psi_2|^2$, and $|\psi_3|^2$
for the minimal vortex in the $U(3)$-symmetric case,
respectively. The energy density is shown in panel (d).
We choose $\hbar=m=1,g = \tilde g=1000,\mu=100$. 
The Rabi frequencies are chosen as $\omega_{12}= \omega_{23} = \omega_{31} = 0.1$ in the first row, and
$\omega_{23} = 0.05$, $\omega_{12} = \omega_{31} = 0.1$ in the second row.
The sizes of the boundary are $L=-15$ to $L=15$ for both $x$ and $y$.
}
\label{fig:1b}
\end{center}
\end{figure}
Let us finally consider a vortex trimer in the $U(3)$ symmetric case
($g = \tilde g$)
where all of the terms in Eq.~(\ref{eq:gp})  except for the last one 
are invariant under $\vec \psi \to U\vec\psi$ with $U \in U(3)$.
A striking difference from the previous case with $g\neq \tilde g$ can be best seen in the limit where $\omega_{ij}\to0$.
The ground state is degenerate and its order parameter space is
$U(3)/U(2) \simeq S^5$ defined by
$\sum_{i=1}^3 |\psi_i|^2 = \frac{\mu}{g}$.
The first homotopy group of the ground state is trivial, and there are no topologically stable vortices.
When the Rabi frequencies are not zero, the order parameter space becomes $U(1)$ implying
the existence of stable vortex configuration.
Fig.~\ref{fig:1b}  shows numerical solutions for several choice of the
Rabi frequencies. 
We examine two choices for the Rabi frequencies  with $g = \tilde g = 1000$
($\mu = 100$ and $\hbar = m =1$): 
i) $\omega_{12}= \omega_{23} = \omega_{31} = 0.1$, which leads to
the ground state condensation $(|\psi_1|,|\psi_2|,|\psi_3|) = (0.183,0.183,0.183)$, and
ii) $\omega_{23} = 0.05$, $\omega_{12} = \omega_{31} = 0.1$, which leads to
 $(|\psi_1|,|\psi_2|,|\psi_3|) = (0.203,0.171,0.171)$.
Although the profiles of the condensation 
are not axisymmetric, 
the total energy density is universally axisymmetric; see the rightmost panels in Fig.~\ref{fig:1b}.
Unlike the previous case with $g \neq \tilde g$, it is impossible to 
see a partonic nature from the energy density when $g=\tilde g$.
These things are related to the fact that there are no constituent vortices standing alone in the limit $\omega_{ij} \to 0$.
This configuration can be regarded
as a skyrmion in the $\mathbb{C}P^2$ nonlinear sigma model 
with the two-dimensional complex projective space 
$\mathbb{C}P^2 \simeq S^5/S^1 \simeq SU(3)/[SU(2)\times U(1)]$, 
instead of $\mathbb{C}P^1 \simeq S^3/S^1 \simeq SU(2)/U(1) \simeq S^2$ 
for two component BECs \cite{Kasamatsu:2004}. 

\section{Discussion}

Finally, we comment on the possibility of realization in 
experiments. 
In this paper, we have solved the GP equation in the static system by fixing the phase winding 
and the constant density at the boundaries [18]. 
This implies that the vortex trimer is created in rotating BEC's 
in laboratory experiments. 
First, an integer vortex is created as usual by gradually increasing 
the rotation, and then it will be split into the vortex trimer. 
Two component BEC's of different hyperfine states of 
the same atom have been already realized using
the $|1,-1\big>$ and $|2,1\big>$ states \cite{Matthews} 
and the $|2,1\big>$ and  $|2,2\big>$ states \cite{Maddaloni:2000} 
of $^{87}$Rb, respectively. 
Three component system should be possible using a mixture of those 
states of $^{87}$Rb. For that, one needs to use the optical trap 
which has been recently been realized in a two-component system 
\cite{Hamner:2011}.

In this paper, we have investigated the vortex trimer in the three-component
BEC's. 
We expect that the vortex $N$-omers can also be constructed 
in $N$-component BEC's. 
However, for $N \geq 4$, the choice of 
the internal coherent couplings $\omega_{ij}$ becomes nontrivial. 
For instance, for $N=4$, we expect that the choice
$\omega_{12}= \omega_{23}= \omega_{34}= \omega_{41}>0$ 
gives a symmetric tetramer, but we still have 
additional parameters 
$\omega_{13}$ and $\omega_{24}$.  
We will study the $N$-omer elsewhere.

\section*{Acknowledgments}
We would like to thank K.~Kasamatsu and S.~Tojo for useful comments.
This work is supported in part by 
Grant-in Aid for Scientific Research (No. 23740198 and No. 23740226) 
and by the ``Topological Quantum Phenomena'' 
Grant-in Aid for Scientific Research 
on Innovative Areas (No. 23103515)  
from the Ministry of Education, Culture, Sports, Science and Technology 
(MEXT) of Japan.


\end{document}